\documentstyle {article}

\newcommand{\Section}[1]{\section{#1}\setcounter{equation}{0}}

\def\QED{\hbox{\kern 1pt\vrule width 3pt height 7pt}}
\def\d{\delta}

\def\a{\alpha}
\def\ep{\epsilon}
\def\b{\bar}
\def\g{\gamma}
\def\th{\theta}
\def\del{\delta}

\def\be{\begin{equation}}
\def\bea{\begin{eqnarray}}

\def\nn{\nonumber}
\def\ll{\lambda}
\def\l{\label}
\def\ee{\end{equation}}
\def\eea{\end{eqnarray}}
\def\C{\rm {I\kern-.520em C}}

\def\D{\partial}
\def\w{WZW }
\def\o{\over}
\begin{document}
\begin{titlepage}
\hfill
\vbox{
    \halign{#\hfil         \cr
           hep-th/9607084 \cr
           IPM-96-128   \cr
           } % end of \halign
      }  % end of \vbox
\vspace*{3mm}
\begin{center}
{\large \bf  GAUGING OF LORENTZ GROUP WZW MODEL BY ITS NULL SUBGROUP }
\vskip .5in
{\bf Amir Masoud Ghezelbash} \footnote{e-mail address:
amasoud@physics.ipm.ac.ir}\\
\vskip .25in
{\em
Institute for Studies in Theoretical Physics and Mathematics, \\
P.O. Box 19395-5531, Tehran, Iran.}\\
{\em  and}\\
{\em
Department of Physics, Alzahra University, \\
Tehran 19834, Iran.}\\
\vskip .5in
\end{center}
\begin{abstract}
We consider the standard vector gauging of Lorentz group $ SO(3,1) $ WZW model
by
its non-semisimple null Euclidean subgroup in two dimensions $ E(2) $.
The
resultant effective action of the theory is seen to describe a one dimensional
bosonic field in the
presence of external charge that we interpret it as a Liouville field.
Gauging a boosted $ SO(3) $ subgroup, we find that
in the limit of the large boost, the theory can be interpreted as an
interacting
Toda theory. We also take the generalized non-standard bilinear form for $
SO(3,1) $ and gauge both $ SO(3) $ and $E(2)$ subgroups and discuss the
the resultant theories.
\end{abstract}
\end{titlepage}
\newpage
\def \dbar {\bar \partial}
\def \d  {\partial}
\Section{Introduction}

The appearance of black hole geometry
in the context of gauged \w models was first discovered in \cite{WITTEN} and
has lead to an
extensive study of gauged \w for different groups and subgroups
\cite{HORN,FRAD}.
In particular in reference
\cite{FRAD} the standard vector gauging of the Lorentz group \w model by its
subgroups
$ SO(3) $ and $ SO(2,1) $ was considered and their singularity structure was
studied.
These two different subgroups belong to
two distinct conjugacy classes of the subgroups of Lorentz group. However,
there is
another conjugacy class of subgroup, that is the subgroup which is isomorphic
to the Euclidean
group in two dimensions; $ E(2) $.
In a previous articles \cite{ARDALAN,ARDALAN2} on $SL(2,R)/U(1)$ black hole,
it was discovered that taking the corresponding Euclidean subgroup,
result an unexpected reduction of the quotient theory
and yield a single Liouville field. In \cite{ARDALAN3}, we studied this
reduction of
degrees of freedom and found out that it was the consequence of a large
symmetry
.\newline
In this article, we consider the target space of Lorentz group \w model
gauged
by its vector non-semisimple null
subgroup $ E(2) $. The resultant target space and dilaton field will be
one dimensional again, but when we boost the $ SO(3) $ subgroup which tends
to $ E(2)$ in the limit of infinite boost, there appears an interacting Toda
structure.
In section two, we consider $ SO(3,1)/SO(3) $ in a
suitable gauge, and find the target space metric and the dilaton field
which
although turn out to be rather
complicated.
In section three, $ SO(3,1)/E(2) $ is presented, and is found that it
describes a one dimensional
Liouville theory.
In section four, $ SO(3,1)/SO(3)_b $ is discussed,
where
$ SO(3)_b $ is the boosted version of $ SO(3) $ group. We will show that in
the
limit $ b \rightarrow \infty $, as expected, the results are the same as
the vector gauged $
SO(3,1)/E(2) $ model,
and for finite but large boost paremeter, an interacting Toda theory appears.
In section five, we investigate the generalized non-standard
bilinear form of Lorentz group in the construction of gauged models.\newline
Finally, in the appendix we fill out the details of our gauge fixing.
\Section{VECTOR GAUGING OF SO(3,1)/SO(3) WZW MODEL}

In this section we consider the $ SO(3,1) $ WZW model vector gauged by its $
SO(3) $ subgroup. We call the six generators of $ SO(3,1) $ as $
J_1,J_2,J_3,K_1,K_2,K_3 $.
The gauged action can be written as
\be \l{GAUGE} S(g,A,\bar A)=S(g)+S_{gauge}(g)=S(g)+{k\over 2\pi}\,\int
_{\Sigma} \,d^2z\,
Tr(-\,\bar AJ+A\bar J+A\bar A-g\bar Ag^{-1}A\,),
\ee
where the gauge fields $ A,\bar A $ takes their values in $ so(3) $
algebra \be (A,\bar A)=(A_1,\bar A_1)\,J_1+(A_2,\bar A_2)\,J_2+(A_3,
\bar A_3)\,J_3.\ee
$ S(g) $ is the ungauged $ SO(3,1) $ WZW model and $ J,\bar J $ are the
corrsponding Kac-Moody currents
\be
S(g)={k\over 4\pi}\,\int
_{\Sigma}\,d^2z\,
Tr(\,g^{-1}\partial gg^{-1} \bar \partial
g\,)-{k\over 12\pi}\,\int _{M}\,
Tr(\,g^{-1}dg\,)^3,
\ee
\be               \l{JJJ}
J=g^{-1}\partial g,\bar J=\bar
\partial gg^{-1}.
\ee
This action is invariant
under following vector gauge transformation
\be g \rightarrow h^{-1}g\,h,A \rightarrow h^{-1}\,(A-\partial)\,h,\bar A
\rightarrow h^{-1}\,(\bar A-\bar \partial)\,h, \ee
$$  g \in SO(3,1),h \in SO(3). $$
By using the homomorphism between $ SL(2,C) $ and $ SO(3,1) $ groups, we
fix the
gauge and obtain ( the details will be discussed in the appendix )
\be \l{GFIX}
g= \pmatrix { r^2+{{1+r^4-2r^2 \cos(2\theta)} \over {2t^2}}+{{t^2} \over
{2}} & r \cos(\theta) ({{-1+r^2+t^2} \over t}) & - r
\sin(\theta)
({{-1+r^2+t^2} \over t}) & {{1+r^4-2 r^2 \cos(2\theta)} \over {2t^2}}-
{{t^2}\over 2} \cr
r \cos(\theta) ({{-1+r^2+t^2} \over t}) & -1+2 r^2 \cos^2(\theta)
& -r^2 \sin(2\theta) & r \cos(\theta) ({{-1+r^2-t^2} \over {t}}) \cr
-r\sin(\theta)
({{1+r^2+t^2} \over t}) & -r^2 \sin(2\theta) & 1+2 r^2 \sin^2(\theta)
& r\sin(\theta) (-{{1+r^2+t^2} \over t}) \cr
-{{1+r^4-2r^2\cos^2(\theta)} \over {2t^2}}+{{t^2} \over 2}
 & r \cos(\theta) ({{1-r^2+t^2} \over t}) & - r
\sin(\theta)
({{-1-r^2+t^2} \over t})
& r^2-{{1+r^4-2r^2\cos(2\theta)} \over {2t^2}}-{t^2 \over 2}} \ee
In this parametrization, the ungauged WZW model action $ S(g) $ becomes
\bea  \l{SG}
S(g)={{2k} \over \pi}\,\int d^2z\,&\{&{{\dbar t\,\d t(1-r^2
\cos(2\theta)} \over {t^2}}-\dbar r\,\d r\,\cos(2\,\theta)+\dbar \theta\,\d
\theta\,r^2\,\cos(2\,\theta) \nonumber \\
&+&(\dbar t\,\d r+\dbar r\,
\d t)\,{{r\,\cos(2\,\theta)} \over t}+(\dbar \theta\,\d r+\dbar r\,\d
\theta)\,r\sin(2\,\theta) \nonumber \\
&-&(\dbar \theta\,\d t+\dbar t\,\d
\theta)\,{{r^2 \sin(2\,\theta)} \over t}\}.
\eea
Note that in our gauge the WZ term in $ S(g) $ vanishes. The
currents (\ref{JJJ}) are given by
\bea  \l{JSJSJS}
J_1 &=& ({{r \cos(\theta)+r^3 \cos(3\,\theta)} \over
t}-r\,t\,\cos(\theta))\,\d \theta+(r \sin(\theta)-r^3
\sin(3\,\theta)+{{r\,\sin(\theta)} \over {t^2}})\,\d t \nonumber \\
&+&({{\sin(\theta)+r^2\,\sin(3\,\theta)} \over t}-t \sin(\theta))\,\d r,
\nonumber \\
J_2 &=& ({{r \sin(\theta)-r^3 \sin(3\,\theta)} \over
{t}}-r\,t\,\sin(\theta))\,\d \theta+(-r \cos(\theta)+
{{r\,\cos(\theta)-r^3 \cos(3\,\theta)} \over {t^2}})\,\d t \nonumber \\
 &+&
({{\cos(\theta)+r^2\,\cos(3\,\theta)} \over {t}}+t \cos(\theta))\,\d r,
\nonumber
\\
 J_3 &=& -2\,r^2\,\cos(2\,\theta)\,\d \theta+{{2 r^2 \sin(2\,\theta)} \over
{t}}\, \d t-2\,r \sin(2\,\theta)\,\d r, \nonumber \\
 J_4 &=& ({{-r \sin(\theta)-r^3 \sin(3\,\theta)} \over
{t}}+r\,t\,\sin(\theta))\,\d \theta+(r \cos(\theta)+{{r\,\cos(\theta)-r^3
\cos(3\,\theta)} \over {t^2}})\,\d t \nonumber \\
 &+&
({{\cos(\theta)+r^2\,\cos(3\,\theta)} \over {t}}-t \cos(\theta))\,\d r,
\nonumber
\\
 J_5 &=&({{-r \cos(\theta)-r^3 \cos(3\,\theta)} \over
{t}}-r\,t\,\cos(\theta))\,\d \theta+(r \sin(\theta)+{{-r
\sin(\theta)+r^3\,\sin(3\,\theta)} \over {t^2}})\,\d t \nonumber \\
 &+&
({{\sin(\theta)-r^2\,\sin(3\,\theta)} \over {t}}-t \sin(\theta))\,\d r,
\nonumber
\\
 J_6 &=& 2\,r^2\,\sin(2\,\theta)\,\d \theta+2({{r^2 \cos(2\,\theta)-1}
\over {t}})\,\d t-2\,r \cos(2\,\theta)\,\d r,
\eea
and similar expressions for $\bar J$'s are obtained from above expressions
for $J$'s with $ \partial $ replaced by $ \dbar $ and a overall minus sign
multiplying in $ J_3,J_6 $ . Integrating out the gauge fields, we obtain the
following result for $ S_{gauge}(g) $
\bea \l{FFFFFF}
S_{gauge}(g)&=&(2k)/(\pi) \int d^2 z \{ H_{tt}\dbar t \d t+H_{rr} \dbar r\d
r+H_{\theta \theta} \dbar \theta \d \theta \nonumber \\ &+&H_{tr} (\dbar t \d
r+\dbar r \d
t)
+H_{\theta r}(\dbar \theta \d r+\dbar r \d \theta)+ H_{\theta
t}(\dbar \theta \d t+\dbar t \d \theta)\},
\eea
where $ H $'s are defined by
$$ H_{ij}=\sum_{k=1}^3 \bar J_{k_i}
\Delta_{k_j}  \qquad (i,j=\theta,t,r). $$
Here $ \bar J_{k_i} $, $ \Delta_{k_
j} $ are the coefficints of $ \bar \d i
$ and $ \d j $ in the $ \bar J_k $ and
$ \Delta_k $ respectively and
\bea
\Delta_1&=&(1/M) \big( r t
\cos (\theta) \{ -1+5 r^4+3 r^2+r^6+2
r^2 t^2+3 r^4 t^2+3 t^2-3
t^4+t^6 \nonumber \\ &+&(r^2+t^2) (3 r^2 t^2-8 r^2 \cos
^2 (\theta)) \}\,\d \theta \nonumber \\
&+&r \sin
(\theta) \{ -1+r^2+r^4-r^6+t^2-2 r^2
t^2+r^4 t^2+t^4+r^2 t^4-t^6-4 r^2 \sin
^2 (\theta) (1+r^2-t^2) \}\,\d t \nonumber \\
&+&
t \sin (\theta) \{ -1-r^2+r^4+r^6+3
t^2-r^2 t^2 (r^2+t^2-6+8 \sin^2 (\theta))-3 t^4+t^6 \}\,\d r \big), 
\nonumber \\
\Delta_2&=&(1/M) \big(r t
\sin (\theta) \{ -1+5 r^4-3 r^2-r^6+2
r^2 t^2-3 r^4 t^2-3 t^2-3
t^4-t^6-8 r^2 \sin
^2 (\theta) (r^2+t^2) \}\,\d \theta \nonumber \\
&+&r \cos
(\theta) \{ 1+r^2-r^4-r^6+t^2+2 r^2
t^2+r^4 t^2-t^4+r^2 t^4-t^6-4 r^2 \cos
^2 (\theta) (1-r^2+t^2) \}\,\d t \nonumber \\
&+&
t \cos (\theta) \{ 1-r^2-r^4+r^6+3
t^2-r^2 t^2 (r^2+t^2-6+8 \cos^2 (\theta))+3 t^4+t^6 \}\,\d r \big), 
\nonumber \\
\Delta_3&=&({1 \over M^{1/2}}) \big(r
\{ -r^3-r t^2
+ r \cos
 (2 \theta) \}\,\d \theta + r \sin
(2 \theta) \d r \big), \nonumber \\
M&=&(1+r^4-t^4-2 r^2 \cos (2 \theta))^2.
\eea
The resulting effective action describes a three dimensional target space
metric
that has singularities in  $ t=0 $ and $ M=0 $. As a consequence of the
conformal invariance of the theory, the dilaton field must
solves the equation $ R_{\mu \nu}=\cal {D}_{\mu} \cal {D}_{\nu} $ $\Phi $,
and is found to be
\be \Phi=2
\ln ({M \over {2 t^4}})+\alpha ,\ee
where $ \alpha $ is a constant. Moreover the
antisymmetric tensor
is zero.\newline
This background metric and dilaton field were obtained previously in a simple
form in reference \cite{FRAD}, with a different gauge fixing of the $ g $
field ( Appendix ).
\Section{VECTOR GAUGING OF SO(3,1)/E(2) WZW MODEL}

In this section, we consider the gauging of the
\w
model by its Euclidean subgroup $ E(2) $.
The generators of this subgroup are
\be E_1=J_1-K_2,
E_2=J_2+K_1,
E_3=J_3.
\ee
Obviously in the four dimensional fundamental representation, the two
generators $ E_1, E_2
$ (generators of translational motions in plane) are null
(and also nilpotent)
\be Tr(E_1^2)=Tr(E_2^2)=E_1^3=E_2^3=0. \ee
As $ E(2) $ is a non-abelian subgroup, the only
consistent gauged WZW model is the vector gauging
\cite{BARS,BARS2}. The gauged action is the same as (\ref{GAUGE})
with the gauge fields $ A,\b A$ taking values in $e(2)$ algebra
\be (A,\bar A)=(A_1,\bar A_1)\,E_1+(A_2,\bar A_2)\,E_2+(A_3,
\bar A_3)\,E_3, \ee
By using the homomorphism between $ SL(2,C) $ and $ SO(3,1) $ groups, we
fix the
gauge ( Appendix ) and obtain the same result for g as (\ref{GFIX}).
In this parametrization, the ungauged WZW model action $ S(g) $ and
the currents (\ref{JJJ}) are like (\ref{SG}), (\ref{JSJSJS}).
Integrating out the gauge fields, we obtain
the following expression for $ S_{gauge}(g) $
\bea \l{SGE2}
S_{gauge}(g)&=&{{2k} \over \pi}\,\int d^2z\,\{{{\dbar t\,\d
t\,r^2
\cos(2\theta)} \over {t^2}}+\dbar r\,\d r\,\cos(2\,\theta)-\dbar \theta\,\d
\theta\,r^2\,\cos(2\,\theta) \nonumber \\
 &-&(\dbar t\,\d r+\dbar r\,
\d t)\,{{r\,\cos(2\,\theta)} \over t}-(\dbar \theta\,\d r+\dbar r\,\d
\theta)\,r\sin(2\,\theta) \nonumber \\
 &+&(\dbar \theta\,\d t+\dbar t\,\d \theta)\,{{r^2
\sin(2\,\theta)} \over t}\},
\eea
From the above relations (\ref{SG}), (\ref{SGE2}) we have
\be
S(g,A,\bar A)={{2k} \over \pi}\,\int d^2z\,({{\dbar t\,\d t
} \over {t^2}}),
\ee
which corresponds to the target space metric \be d\,s^2=2\,k{{d\,t^2}
\over {t^2}} \ee and the dilaton field is found to be
\be \Phi=4 \ln t+\Phi_0, \ee
where $ \Phi_0 $ is a constant. The interesting feature of this result is
that
two degrees of the coset manifold $ SO(3,1)/E(2) $ has disappeared,
reminiscent
of the $
SL(2,R)/E(1) $ results \cite{ARDALAN}. The effective target space action
in conformal
gauge in terms of $
t^2=e^w $ reads as follows
\be \l{SEFFE}
S_{eff}={{k} \over {2 \pi}}\,\int d^2z\, \d w\,\dbar w+{1 \over {4 \pi}}
\,\int d^2z\, \sqrt h\,R^{(2)}\, w,
\ee
which is the action of Liouville theory
without cosmological constant.
As discussed in \cite{ARDALAN3}, this dimensional reduction
is
related to the equivalence of the vector and chiral gauged \w
models when gauged by null subgroups. In fact, in the context of
chiral gauging, by using
the right subgroup as $ H G_{+} $ where $ G_{+} $ is generated by
$ E_1,E_2 $,
and $ H $ by $ E_3 $;
and the left
subgroup $ G_- H $ where $ G_- $ is generated by $
E_{-1}=J_1+K_2 $ and $E_{-2}=- J_2+K_1 $, five degrees of freedom of the
$ g $ field can
be fixed and the resultant effective action again becomes one dimensional.
\Section{VECTOR GAUGING OF $ SO(3,1)/SO(3)_b $ WZW MODEL}

In the previous section, we noted that the target space obtained by gauging
the
$ SO(3,1) $ WZW model by its null subgroup $ E(2) $, degenerated to a
one
dimensional flat space. We know that the group $ SO(3)_b $, obtained by
boosting
the generators of $ SO(3) $ group, tends
to $ E(2) $ group
in the limit of infinite boost parameter.
To understand the mechanism of this two dimensional
reduction, we gauge the $ SO(3,1) $ WZW model by its boosted subgroup
$ SO(3)_b $ and we will see that in the limit of the infinite boost the
target
space metric and the dilaton field tend to the previous result of the
$ SO(3,1)/E(2) $, and for finite but large boost a Toda structure appears.
The $ {SO(3)}_b $ group is generated by
\bea
{(J_1)}_b &=&\exp ({-{b \over 2}K_3})\,J_1\,\exp ({{b \over 2}K_3}) \nonumber
\\ {(J_2)}_b &=&\exp ({-{b \over 2}K_3})\,J_2\,\exp ({{b \over 2}K_3})
\nonumber
\\ {(J_3)}_b &=& J_3.
\eea
We fix the gauge freedom and obtain the $ g $ field ( Appendix ) in the
form of(\ref{GFIX}) by replacing every $t$ by $te^{b}$.
The $ S(g) $ part of $ S(g,A,\bar A) $ is the same as (\ref{SG}). The
currents (\ref{JJJ})
are the same as (\ref{JSJSJS}) with the difference that every $ t $
must be replaced by
$ t\,e^{b} $. The gauge fields are
\be {(A,\bar A)=(A_1,\bar
A_1)\,(J_1)_b+(A_2,\bar A_2)\,(J_2)_b+(A_3,\bar A_3)\,J_3}. \ee
Integrating them out,
we obtain
\bea
S_{gauge}(g)&=&{{2k} \over \pi}\,\int d^2z\,[{{\dbar t\,\d
t\,r^2
\cos(2\theta)} \over {t^2}}+\dbar r\,\d
r\,(\cos(2\,\theta)+{{4 e^{-2b}} \over {t^2}}) \nonumber \\
&-&\dbar \theta\,\d\theta\,(r^2\,\cos(2\,\theta)+{{4 r^2 e^{-2b}} \over
{t^2}})
-(\dbar t\,\d r+\dbar r\,
\d t)\,({{r\,\cos(2\,\theta)} \over t}-{{2 r e^{-2b}}
\over {t^3}}) \nonumber \\ &-&(\dbar \theta\,\d r+\dbar r\,\d
\theta)\,r\sin(2\,\theta )+
(\dbar \theta\,\d t+\dbar t\,\d \theta)\,{{r^2
\sin(2\,\theta)} \over t}+ O(e^{-4b})\,].
\eea
Then the action of $ SO(3,1)/{SO(3)}_b $ reads
\be   \l{PO}
S(g,A,\bar A)={{2k} \over \pi}\,\int d^2z\,[{\dbar t\,\d
t \over {t^2}}+e^{-2b}({{4 \dbar r\,\d r} \over {t^2}}
-{{2\,r(\dbar t\,\d r+\dbar r\,
\d t)} \over {t^3}}+4 r^2 {{\dbar \theta\,\d \theta}
\over {t^2}})+ O(e^{-4b})\,],
\ee
and the dilaton field is found to be
\be \Phi=4 \ln t+\Phi_0+O(e^{-4b}).\ee To diagonalize (\ref{PO}),
we define
\be r^2=e^{V},t=(e^{W}+2e^{-2b+V})^{1/2},\theta=X/2, \ee
and get
\bea    \l{LLL}
S_{eff}&=&{{k} \over {2 \pi}}\int d^2\,z \,\{{{\dbar W \d W}}+
4 \ep ^2 e^{-W+V}( -\dbar W \d W+\dbar X \d X+\dbar V \d V )+O(\ep ^{4}) \}
\nonumber
\\
&+&{1 \over {4 \pi}}\int d^2\,z \, \sqrt h\, R_{(2)}\, \{W+2 \ep ^2e^{-W+V}+
O(\ep ^{4}) \},
\eea
where $\ep =e^{-b}$. We note that in the infinite boost limit,
$\ep \rightarrow 0 $, the above effective action
tends to the previous result for $ SO(3,1)/E(2) $ model. For small $\ep $
, the effective action is the effective action of Toda field $ W $
that interacts with the matter field $ X, V $. Elliminating $X$ field by its
equation of motion from (\ref{LLL}), we find
\bea  \l{LLLL}
S_{eff}&=&{{k} \over {2 \pi}}\int d^2\,z \,\{{{\dbar W \d W}}+
4 \ep ^2 f e^{W-V}+4\ep ^2e^{V-W}( -\dbar W \d W+\dbar V \d V )+O(\ep ^{4}) \}
\nonumber
\\
&+&{1 \over {4 \pi}}\int d^2\,z \, \sqrt h\, R_{(2)}\, \{W+2 \ep ^2e^{-W+V}+
O(\ep ^{4}) \},
\eea
where f is an arbitrary constant. The first term shows the Toda structure
for $W$
field, and the absence of kinetic term for the other Toda field, the $V$
field,
is the result of the existence of just one Cartan generator $J_3$ in the
gauged
ubalgebra. The second term in (\ref{LLLL}) is in the form of a potential
term in Toda
theory. The other terms describe the interaction of the Toda fields $W$
and $V$. Here
we have a transition at $b=0$ from the black hole metric given by (\ref{SG}),
(\ref{FFFFFF}) to a Toda structure at $b \rightarrow \infty$, and between
these two limits
, the effects of black hole structure appear in the form of interaction terms
of the Toda fields.
\Section{VECTOR GAUGING OF $ SO(3,1) $ WZW MODEL BY ITS GENERALIZED
NON-STANDARD BILINEAR FORM}

It is known that the bilinear form that is used in construction of WZW
models,
must be non-degenerate and several interesting models with a bilinear form
different
from the Killing form have been constructed \cite{NAPPI}. In this article we
have
used the usual bilinear form of Lorentz group i.e. its Killing form
\be \l{KILLING}
\Omega=2\{\sum _{i=1}^3 (-e_{ii}+e_{(i+3)(i+3)})\}.
\ee
However, there is also a second non-degenerate bilinear form that can be used
for
$SO(3,1)$, i.e.
\be \l{KILLING2}
\Omega '=\sum _{i=1}^3 \{ e_{i(i+3)}+e_{(i+3)i}\}.
\ee
We can use a linear combination of the two  bilinear forms (\ref{KILLING})
and
(\ref{KILLING2})
\be
\Omega (\ll)={{\Omega} \o {2}}+\ll \Omega ',
\ee
in our construction.\newline
The ungauged WZW action is
\be
S(g)={k \o {4\pi}}\int d^2\,z J_i\Omega _{ij}(\ll){\tilde J}_j,
\ee
where ${\tilde J}_i$ are obtained from (\ref{JSJSJS}) by $\d \rightarrow \bar
\D $
. The gauged part of (\ref{GAUGE}), is
\be
S_{gauge}(g)={k \o {2\pi}}\int d^2\,z \{A_i \b J_j-\b A_iJ_j+A_i\b A_j-
(g^{-1}Ag)_i
\b A_j\}\Omega _{ij}(\ll).
\ee
After integrating out the gauge fields, the target space metric for $ SO(3)$
subgroup becomes
\be \l{DS2}
ds^2={k\o \pi}G_{ij}(\ll)dx^idx^j,\quad x^i=t,\theta,\phi
\ee
where
\bea
G_{tt}(\ll)&=&{2 \o {1+\ll ^2}}+2{\ll ^2 \o {1+\ll ^2}}[{{\ll ^2} \o {\cos ^2
\phi \sin ^2 \theta}}-4{{\ll \tan \phi \cot \theta} \o {\sinh 2t \sin \theta
}}
\nn \\
&+&{{\cosh ^4t-\cos ^2\phi \cosh ^2t-\cos ^2 \th \cos ^2\phi \sinh ^2t+
\sinh ^2
t\cos ^2 \phi \sin ^2 \th \cosh ^2t} \o {\sinh ^2 t \cosh ^2 t \sin ^2 \th
\cos ^2 \phi}}
],\nn \\
G_{\th \th}&=&{2 \o {1+\ll ^2}} \{\coth ^2 t+\ll ^2\},
\nn \\
G_{\phi \phi}&=&{2 \o {1+\ll ^2}} \{ {{\tanh ^2 t} \o {\sin ^2 \th}}+
\cot ^2 \th \coth ^2t\tan ^2\phi-4{{\ll \tan \phi\cot \th} \o {\sinh 2t\cos
\th}}
+\ll ^2(1+{{\cot ^2 \th}\o {\cos ^2\phi}})\},
\nn \\
G_{t \th}&=&
G_{\th t}={{2\ll} \o {1+\ll ^2}} \{-{{\ll ^2\tan \phi} \o {\sin \th}}+
2{{\ll \cot \th} \o {\sinh 2t}}-\coth ^2t {{\tan \phi} \o {\sin \th}}\},
\nn \\
G_{t \phi}&=&
G_{\phi t}={{2\ll} \o {1+\ll ^2}} \{ \ll ^2 \cos \th (\sin ^2 \th \cosh ^2t-{
1 \o {\sin ^2\th\cos ^2\phi}})+2\ll\tan \phi{{1+\cos ^2\th} \o {\sin ^2\th
\sinh 2t}}
\nn \\
&+&4\cos \th ({{-\cosh ^4t(1+\cos ^2 \phi\sin ^2\th)+\cos ^2\phi(-1+\sin
^2\th\cosh ^2t+2\cosh^2t)}\o {\sinh ^22t\cos ^2\phi\sin^2\th}}
\nn \\
&+&{{1+\cosh ^2t\sin ^2\th} \o {2(1+\ll ^2)}})],
\nn \\
G_{\th \phi}&=&
G_{\phi \th}={{2} \o {1+\ll ^2}} \{ \ll ^2 \cot \th \tan \phi-{{2\ll} \o {
\sin \th \sinh 2t}}+\cot \th \tan \phi \cosh ^2 t\},
\eea
and the dilaton field becomes
\be
\Phi =\ln \{ 2\sin ^2\th \sinh ^2 2t \cos ^2 \phi (1+\ll ^2)\}.
\ee
The central charge of coset model is
\be
c=6{{k(k-4)+\ll ^2}\o {(k-4)^2+\ll ^2}}-{{3k} \o {k+2}},
\ee
that by the following redefinition of $k$,
\be
k'={{D+4M} \o {M-D}}\pm {{1} \o {M-D}}\sqrt {9D^2+16M^2}
\ee
where
\bea
D&=&k^3+4k^2+(\ll ^2-32)k+4\ll^2,\nn \\
M&=&
k^3-6k^2+\ll ^2k+2\ll ^2+32
\eea
could be written as
\be
c=3{{k'^2+8k'} \o {(k'+2)(k'-4)}}
,
\ee
that is the usual central charge of $SO(3,1)/SO(3)$ coset model.\newline
If we set $\ll =0$ we recover the target space metric and the dilaton field
that
obtained in reference \cite{FRAD}. In addition, we find that the singularity
structure
of (\ref{DS2}) is like the case with
$\ll =0$;
therefore the singularity structure  doesn't change when using the generalized
bilinear form in place of the usual Killing form of the Lorentz group.\newline
In the case of gauging the $E(2)$ subgroup, the resultant effective action is
in
the form (\ref{SEFFE}) with an overall numerical factor $1+\ll ^2$. This
simple
renormalization of the action, has been independently noted in a study of the
non-standard $SO(3,1) $ WZW model \cite{PARVIZI}.

\Section{CONCLUSIONS}

By studying the structure of the vector $ G/H $ \w model, when  $ G $ is the
Lorentz group
and $ H $ the non-semisimple null Euclidean subgroup in two dimensions $ E(2) $,
we noticed that only one dimension of $ SO(3,1)/E(2) $ three dimensional coset
manifold
survives in the target space metric and the dilaton field of the gauged theory
.
To see the disappearance of the two dimensions from the target space and
dilaton field,
we studied the $ SO(3,1)/SO(3) $ vector gauged model that has a rich
singularity
structure in its target metric. By boosting the $ SO(3) $ gauged subgroup,
at the
limit of infinite boost, the singularity structure disappears and the
three dimensional target space metric reduces to our one dimensional flat
target space.
\newline
This dimensional reduction, on the other hand can be related to the
equivalence
of vector and chiral gauged Lorentz \w models by appropriate null
subgroups \cite{ARDALAN3}.\newline It must be noticed that in \cite{KLIM},
the
gauging of maximally
non-compact groups based on the Gauss decomposition, by some of their null
subgroups was considered. In
particular the null
gauging of non-compact $ SO(3,1) $ has also considered; however, their
gauging
is different from what we consider here. They considered the
gauging of $ SO(3,1) $ by one left and
one right null subgroup which are generated
by a pair of null generators and
the resultant target space metric is different from the one considered here
for
vector gauging of $ E(2) $ subgroup. In particular, the background metric is
four dimensional and no dimensional reduction occurs.\newline
All the above considerations can be applied to the $
SO(n,1)/E(n-1) $ vector \w models, in which cases we also obtain a one
dimensional target space and a dilaton field that can be interpreted as a
Liouville field with a background charge. The $ SO(n,1)/ SO(n)_b $ vector \w
models also tend to the corresponding
$ SO(n,1)/E(n-1) $ models in the limit of infinite boost; and in the finite
but large limit of boost parameter,
the interactions of Liouville field with the other remaining degrees of
freedom occur, giving Toda like theories.

\Section{APPENDIX}

For every element of $ SL(2,C) $ group in the parametrization form $ \pmatrix
{\a & \beta \cr \g & \del} $, there is a $ 4\times4 $ matrix in the
fundamental
representation of $ SO(3,1) $ given by \cite{NAIR},
\be  \l{AKH}
\pmatrix {
{1 \over 2}({\a \a ^*+\beta\beta^*+\g \g ^*+\del \del ^*}) & \Re (\a \beta^*
+\g \del
^*) & - \Im (\a \beta^*+\g \del ^*) &  {1 \over 2}({\a \a ^*-\beta\beta^*+
\g \g
^*-\del \del ^*}) \cr \Re (\a \g ^*+\beta\del ^*) & \Re (\a \del ^*+\beta\g ^*)
&
-\Im (\a \del ^*-\beta\g ^*) & \Re (\a \g ^*-\beta\del ^*) \cr
\Im (\a \g ^*+\beta\del ^*) & \Im (\a \del ^*+\beta\g ^*) & \Re (\a \del
^*-\beta\g ^*) & \Im (\a \g ^*-\beta\del ^*) \cr
{1 \over 2}({\a \a ^*+\beta\beta^*-\g \g ^*-\del \del ^*}) & \Re (\a \beta
^*-\g \del ^*) & -\Im (\a \beta^*-\g \del ^*) & {1 \over 2}({\a \a
^*-\beta\beta^*-\g \g ^*+\del \del ^*}) }
\ee
Let us take the parametrization $ \pmatrix {e^{i \lambda} & 0 \cr \nu
& e^{-i \lambda}} ; 0 \leq \lambda \leq 2\pi,\nu \in C $
for the group elements of $ E(2) $ \cite{MACK}, then by using the vector gauge
freedom, every element of $ SL(2,C)/E(2) $ can be written
as $ \pmatrix {re^{i \theta} & t \cr {{-1+r^2e^{2 i \theta}} \over t} &
re^{i \theta}} $ and by using the above relation (\ref{AKH}) we obtain the
equation
(\ref{GFIX}) for gauged fixed $ g $ field. Similarly for the $ SO(3) $ case,
we
obtain also (\ref{GFIX}).
The gauged fixed field $ g $ that was used in reference
\cite{FRAD}, written in our $ SL(2,C) $ language, is
$ \pmatrix { A+i B & i C \cr i C & {{1-C^2} \over {A+i B}} } $, where $ A,B,C
$ are real fields.

\vskip .5in

\vskip .5in

\noindent{\bf{Acknowledgement}}
\vskip .1 in
I would like to thank F. Ardalan, H. Arfaei
and S. Parvizi for useful
discussions.

\end{document}